\documentclass[final,5p,times,twocolumn,fleqn]{elsarticle}

\usepackage{epsfig}
\usepackage{here}
\usepackage{amssymb}

\usepackage{lineno}

\usepackage{color}
\usepackage{amsmath}

\journal{Physics Letters B}

\newcommand{\km}{\ensuremath{\mathrm{km}}}

\begin{document}

\begin{frontmatter}



\title{Search for sub-GeV dark matter by annual modulation using XMASS-I detector}
\author[ICRR]{M.~Kobayashi\fnref{MKobaNow}}
\author[ICRR,IPMU]{K.~Abe}
\author[ICRR,IPMU]{K.~Hiraide}
\author[ICRR,IPMU]{K.~Ichimura}
\author[ICRR,IPMU]{Y.~Kishimoto}
\author[ICRR,IPMU]{K.~Kobayashi}
\author[ICRR,IPMU]{S.~Moriyama}
\author[ICRR,IPMU]{M.~Nakahata}
\author[ICRR,IPMU]{H.~Ogawa\fnref{OgawaNow}}
\author[ICRR]{K.~Sato}
\author[ICRR,IPMU]{H.~Sekiya}
\author[ICRR]{T.~Suzuki}
\author[ICRR,IPMU]{A.~Takeda}
\author[ICRR]{S.~Tasaka}
\author[ICRR,IPMU]{M.~Yamashita}
\author[ICRR,IPMU]{B.~S.~Yang\fnref{YangNow}}

\author[IBS]{N.~Y.~Kim}
\author[IBS]{Y.~D.~Kim}

\author[ISEE,KMI]{Y.~Itow}
\author[ISEE]{K.~Kanzawa}
\author[ISEE]{K.~Masuda}

\author[IPMU]{K.~Martens}
\author[IPMU]{Y.~Suzuki}
\author[IPMU]{B.~D.~Xu}

\author[Kobe]{K.~Miuchi}
\author[Kobe]{N.~Oka}
\author[Kobe,IPMU]{Y.~Takeuchi}
\author[KRISS,IBS]{Y.~H.~Kim}
\author[KRISS]{K.~B.~Lee}
\author[KRISS]{M.~K.~Lee}
\author[Miyagi]{Y.~Fukuda}
\author[Tokai1]{M.~Miyasaka}
\author[Tokai1]{K.~Nishijima}
\author[Tokushima]{K.~Fushimi}
\author[Tokushima]{G.~Kanzaki}
\author[YNU1]{S.~Nakamura}

\address{\rm\normalsize XMASS Collaboration$^*$}
\cortext[cor1]{{\it E-mail address:} xmass.publications13@km.icrr.u-tokyo.ac.jp .}

\address[ICRR]{Kamioka Observatory, Institute for Cosmic Ray Research, the University of Tokyo, Higashi-Mozumi, Kamioka, Hida, Gifu, 506-1205, Japan}
\address[IBS]{Center for Underground Physics, Institute for Basic Science, 70 Yuseong-daero 1689-gil, Yuseong-gu, Daejeon, 305-811, South Korea}
\address[ISEE]{Institute for Space-Earth Environmental Research, Nagoya University, Nagoya, Aichi 464-8601, Japan}
\address[IPMU]{Kavli Institute for the Physics and Mathematics of the Universe (WPI), the University of Tokyo, Kashiwa, Chiba, 277-8582, Japan}
\address[KMI]{Kobayashi-Maskawa Institute for the Origin of Particles and the Universe, Nagoya University, Furo-cho, Chikusa-ku, Nagoya, Aichi, 464-8602, Japan}
\address[Kobe]{Department of Physics, Kobe University, Kobe, Hyogo 657-8501, Japan}
\address[KRISS]{Korea Research Institute of Standards and Science, Daejeon 305-340, South Korea}
\address[Miyagi]{Department of Physics, Miyagi University of Education, Sendai, Miyagi 980-0845, Japan}
\address[Tokai1]{Department of Physics, Tokai University, Hiratsuka, Kanagawa 259-1292, Japan}
\address[Tokushima]{Department of Physics, Tokushima University, 1-1 Minami Josanjimacho Tokushima city, Tokushima, 770-8506, Japan}
\address[YNU1]{Department of Physics, Faculty of Engineering, Yokohama National University, Yokohama, Kanagawa 240-8501, Japan}

\fntext[MKobaNow]{Now at Physics Department, Columbia University, New York, NY 10027, USA.}
\fntext[OgawaNow]{Now at Department of Physics, College of Science and Technology, Nihon University, Kanda, Chiyoda-ku, Tokyo 101-8308, Japan.}
\fntext[YangNow]{Now at Center for Axion and Precision Physics Research, Institute for Basic Science, Daejeon 34051, South Korea.}

\begin{abstract}

A search for dark matter (DM) with mass in the sub-GeV region (0.32--1 GeV) was conducted by looking for an annual modulation signal in XMASS, a single-phase liquid xenon detector. 
Inelastic nuclear scattering accompanied by bremsstrahlung emission was used to search down to an electron equivalent energy of 1 keV.
The data used had a live time of 2.8 years (3.5 years in calendar time), resulting in a total exposure of 2.38 ton-years. 
No significant modulation signal was observed and 90\% confidence level upper limits of $1.6 \times 10^{-33}$ cm$^2$ at 0.5 GeV was set for the DM-nucleon cross section.
This is the first experimental result of a search for DM mediated by the bremsstrahlung effect.
In addition, a search for DM with mass in the multi-GeV region (4--20 GeV) was conducted with a lower energy threshold than previous analysis of XMASS.
Elastic nuclear scattering was used to search down to a nuclear recoil equivalent energy of 2.3 keV, and upper limits of 2.9 $\times$10$^{-42}$ cm$^2$ at 8 GeV was obtained.

\end{abstract}


\end{frontmatter}


%
%

\section{Introduction}

The nature of dark matter (DM) is a key mystery in cosmology, and detecting it via any force other than gravity is essential for advancing particle physics beyond the standard model.
Weakly interacting massive particles (WIMPs) at $O$(100 GeV) are predicted by theoretical extensions of the standard model, such as the constrained minimal supersymmetric standard model and are strong DM candidates {\cite{WIMPs}.
They have been investigated extensively via nuclear recoil \cite{XENON1T,LUX,PANDA}; however, no significant detections of WIMPs have been confirmed.

Other theories predict a myriad of different DM types, light-mass WIMPs \cite{DUAN2018296}, asymmetric DM \cite{AsymmDM1,AsymmDM2,AsymmDM3}, or hidden sector DM {\cite{hiddenDM} and many others; the mass of these DM candidates ranges from sub-GeV to a few GeV.
Semi-conductor and crystal detectors have searched for these light DM candidates by lowering their nuclear recoil energy thresholds \cite{CDMSLitePaper, CRESSTSurPaper}.
A search via DM-electron scattering by existing detectors have also been performed {\cite{DM_elec, DM_elec_SLAC}.
In addition to these detectors, conventional xenon detectors should also be sensitive to DM with sub-GeV mass \cite{subGeV, subGeV2}, due to the irreducible contribution of the bremsstrahlung effect accompanying nuclear recoils \cite{subGeV}. 
The bremsstrahlung effect can occur when DM collides with a nucleus causing it to recoil and accelerate. 
In the case that a mass of DM particle is 1 GeV, the energy deposited by the bremsstrahlung photon is at most 3 keV. This energy is considerably more than that deposited by elastic nuclear recoil ($\sim$0.1 keV).

In addition to this bremsstrahlung effect, another inelastic effect called the Migdal effect} has also been suggested \cite{Migdal}.
This effect leads to the emission of an electron from the atomic shell and causes subsequent radiation through the inelastic recoil of DM and nuclei.}
Although the bremsstrahlung and Migdal effects need both be calibrated experimentally in xenon and cross sections are smaller than that of elastic nuclear recoil ($\sim$$10^{-6}$ for Migdal, $\sim$$10^{-8}$ for Bremsstrahlung at 1 GeV), 
because these inelastic effects lead to larger energy deposition than elastic nuclear recoil, it should be possible to detect sub-GeV DM through these effects.

Moreover, searching for a spin-dependent (SD) interaction utilising these effects is an attractive possibility, 
since xenon has a larger fraction of odd isotopes than that of other isotopes, such as oxygen \cite{CRESSTSurPaper}. 
Xenon has two stable odd isotopes, namely $^{129}$Xe and $^{131}$Xe which account for 26.4\% and 21.2\% of the natural xenon abundance, respectively; oxygen has only 0.04\% of odd isotopes.
Further theoretical studies are expected to enable the quantitative interpretation of the SD interaction by sub-GeV DM.

This letter reports on the first experimental search for sub-GeV DM (0.32--1.0 GeV) utilizing the bremsstrahlung effect.
In the case of xenon, the Migdal effect is accompanied by M-shell electron emission, and the most likely de-excitation energy is 0.66 keV from the 3d orbit.
As discussed in Section 4, since our understanding of detector responses is limited to those greater than 1 keV, we focus only on for the signal from the bremsstrahlung effect in this analysis.
On the other hand, the search for multi-GeV DM (4--20 GeV) via conventional elastic nuclear recoils \cite{XMASS_MOD, XMASS_MOD2017} was performed. 
For multi-GeV DM search, data with lower energy threshold than in previous studies \cite{XMASS_MOD, XMASS_MOD2017} were used to improve sensitivity in the low mass range.
These searches were conducted by looking for the annual modulation of the event rate in the XMASS data.

\section{Expected annual modulation of signal}

The annual modulation of the bremsstrahlung signal from the sub-GeV DM is evaluated by following the study in \cite{subGeV}.
The differential cross section for such a process is

\begin{eqnarray}
\label{eq:dsigdomg}
 \frac{d\sigma}{d\omega} = 
 \frac{4\alpha |f(\omega)|^2}{3\pi \omega} \frac{\mu_N^2 v^2\sigma_0^{{\rm SI}} }{m_N^2} 
 \sqrt{1-\frac{2\omega}{\mu_N v^2}} \left( 1 - \frac{\omega}{\mu_N v^2}  \right), 
\end{eqnarray}
where $\omega$ is the bremsstrahlung photon energy, $\alpha$ is the fine structure constant, $f(\omega)$ represents atomic scattering factor, $\mu_N$ is the DM-nucleus reduced mass, $v$ = $|{\bf v}|$ is the absolute value of the relative velocity between DM and the target ${\bf v}$, $m_N$ is the nucleus mass, $\mu_N$ is the DM-nucleus reduced mass, $\sigma_0^{SI} \simeq A^2 \sigma_n (\mu_N/\mu_n)^2 $ is the spin-independent DM-nucleus cross section in which $\sigma_n$ is the DM-nucleon elastic cross section, $\mu_n$ is the DM-nucleon reduced mass, and $A$ is the atomic mass number.
The cross section of bremsstrahlung effect is suppressed by the factor of $\frac{\alpha}{m_{N}^2}$ from that of elastic nuclear recoil.

 The corresponding differential event rate is
\begin{equation}
\label{eq:velavg}
 \frac{dR}{d\omega} = N_T \frac{\rho_{\chi}}{m_{\chi}} \int_{ v\geq v_{\rm min}} 
 d^3 v v f_v({\bf v} + {\bf v_E}) \frac{d\sigma}{d\omega} ,
\end{equation}
where $N_T$ is the number of target nuclei per unit mass in the detector, $\rho_{\chi}$ = 0.3 GeV cm$^{-3}$ is the local DM mass density \cite{PDG}, $m_{\chi}$ is the DM mass, ${\bf v_{E}}$ is the velocity of the Earth relative to the galactic rest frame.
$f_v({\bf v})$ is the DM velocity distribution in the galactic frame. 
It is assumed to be a truncated Maxwellian distribution with escape speed $v_{\rm esc } = 544\,\km/\rm{s}$, most-probable velocity $v_0 = 220\,\km/\rm{s}$ and minimum velocity $v_{\rm min}=\sqrt{2\omega/\mu_N}$ \cite{Lewin}.
Assuming that the relative velocity between DM and detector varies as $\{232 + 15\sin2\pi (t - \phi)/T\}$~km/s \cite{MW}, in which the phase $\phi$ = 152.5 days \cite{PDG} from January 1st and period $T$ = 365.24 days, we calculated the event rate as a function of bremsstrahlung energy and time.
Figure \ref{signal} shows the expected bremsstrahlung spectra for 0.5 GeV DM at June and December corresponding to the maximum and minimum $v_{E}$, respectively, as well as the averaged spectrum.
The expected modulation amplitude is about 30\% of the average event rate at 1 keV before considering the effect of the detector such as energy non-linearity or resolution.

The annual modulation in the conventional nuclear recoil signal caused by DM has also been discussed as in \cite{Lewin}.
To evaluate the amplitude for this signal, the same calculation in the previous analysis by XMASS was performed \cite{XMASS_MOD,XMASS_MOD2017}.

\begin{figure}[!t]
  \centering
  \includegraphics[width=8cm]{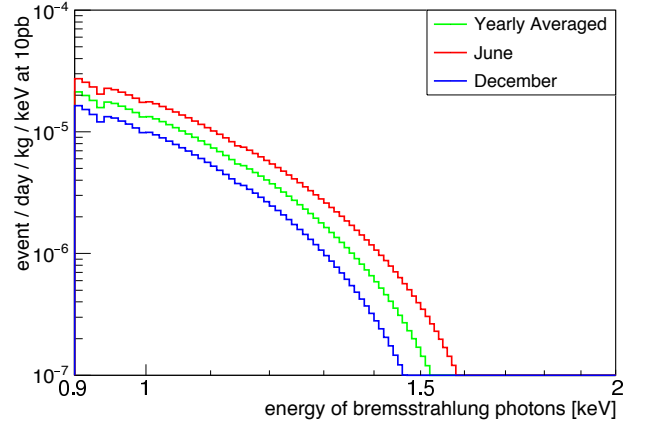}
  \caption{(Colour online) Expected energy spectra of bremsstrahlung caused by 0.5 GeV DM. 
  The red and blue lines represent the spectra in June and in December, respectively, and the green line represents the annual average spectrum the annual average spectrum before considering the effect of detector.}
  \label{signal}
\end{figure}

\section{XMASS Experiment} 

The XMASS-I detector is a single-phase liquid xenon (LXe) detector located underground (2,700 meter water equivalent) at the Kamioka Observatory in Japan \cite{XMASS_Det}.
The inner detector contains 832 kg of xenon and has a pentakis-dodecahedron structure made of copper that supports 642 Hamamatsu R10789 photomultiplier tubes (PMTs).
The quantum efficiency of the R10789 at room temperature is $\sim$30\%. 
The PMTs cover more than 62\% of the inner surface resulting in a large number of photoelectrons per keV detected by the PMTs (PE yield), as it is $\sim$15 PE/keV for 122 keV $\gamma$ ray with zero electric field. 
Here, one PE is defined as the average PE observed at one photon incident to correct for the double PE emission from a PMT in the case of the xenon scintillation \cite{DoublePE}.
Signals from PMTs are recorded by waveform digitizers (CAEN V1751) with 1 GHz sampling rate.
To shield the detector from external neutrons and $\gamma$-rays while also providing a muon veto, XMASS-I sits at the centre of a cylindrical water-Cherenkov detector.
The Cherenkov detector is 10.5 m in height, 10 m in diameter and has 72 Hamamatsu H3600 PMTs arranged on the inside of its wall.

This work used the data collected between November 20, 2013 and June 20, 2017.
The xenon was required to maintain a stable operational temperature and pressure.
A detailed plot of the LXe temperature and pressure during the first 2.7 years of this dataset are shown in \cite{XMASS_MOD2017}, and the values were kept consistently within 0.05 K and 0.2 kPa in the following year.
Periods with the problem of data acquisition system or electronics, such as excessive PMT noise, or unstable pedestal levels were removed from the dataset.
The dataset has a total live time of 2.8 years, and the exposure is 2.38 ton-years.
In addition to this data set, data with a lower energy threshold has also been taken since December 8, 2015.
This data, referred to as low threshold data has 0.63 ton-year of exposure, and is used only for multi-GeV analysis.
Details are discussed in section 6.
In Fig.~\ref{data}, observed data and simulated signal for bremsstrahlung and nuclear recoils are shown. 

\begin{figure}[!t]
  \centering
  \includegraphics[width=9cm]{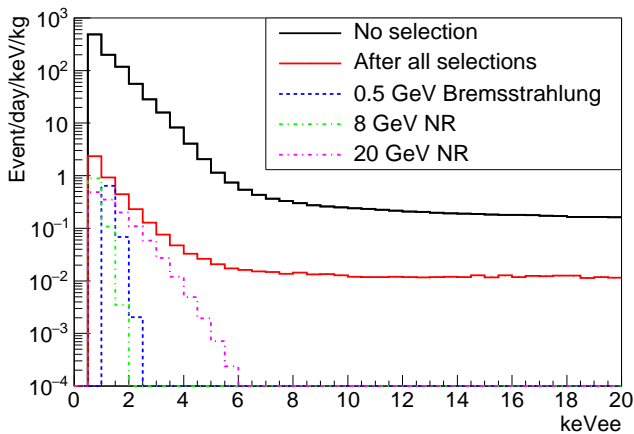}
  \caption{(Colour online) Energy spectra of observed data before selection (solid black line), after selection (solid red line), and of signal simulation after selection (dashed lines). 
  A dashed blue line represents the bremsstrahlung effect from 0.5 GeV with $3 \times 10^{-32}$ cm$^2$ of cross section.
  Dashed green and pink lines represent nuclear recoil from 8, 20 GeV with $10^{-41}$, $10^{-42}$ cm$^2$ of cross section, respectively.
  }
  \label{data}
\end{figure}

\section{Calibration}
The gain of each PMT was monitored by measuring single PE using a blue LED attached to the inner surface of the detector.
This LED is flashed once per second, and gain of each PMT was calculated based on the weekly averaged LED data.
The PE yield was tracked by inserting a $^{57}$Co source into the detector every one or two weeks.
These calibration processes are described in detail in \cite{XMASS_MOD,XMASS_MOD2017,XMASS_Det,XMASS_Cal}.
The PE yield, absorption and scattering length for the scintillation light as well as the number of generated LXe scintillation photons per keV (light yield), are evaluated from the $^{57}$Co calibration data with the help of a Monte Carlo (MC) simulation. 
In the simulation, two PE emissions are also taken into account.
The variation in PE yield can be explained by changes of the absorption length in the LXe \cite{XMASS_MOD2017}. 
To reduce this change of PE yield, xenon gas has been purified continuously by circulating through hot metal getters since March 2015.
The standard deviation of the PE yield was $\pm$2.4\% and $\pm$0.5\% before and after the circulation has been started, respectively.

In this letter, two different energy scales, ``keV$_{\rm{ee}}$'' and ``keV$_{\rm{nr}}$'', are used to indicate the electron-equivalent energy and nuclear recoil energy, respectively.
These are different from those used in the previous analysis \cite{XMASS_MOD,XMASS_MOD2017} below 5.9 keV$_{\rm{ee}}$ and 3 keV$_{\rm{nr}}$ as new calibrations were performed in this low energy region as explained as follow.

For the electron-equivalent energy, the non-linearity of the light yield (scintillation efficiency) along energy was taken into account using the model from Doke et al. \cite{DOKE} with corrections based on the result of calibration. 
The scintillation efficiency below 5.9 keV was calibrated using the L-shell X-ray escape peaks measured during calibration with an $^{55}$Fe source.
These escape peaks distribute energy in 1.2--2 keV, and the weighted mean energy of these escape peaks was 1.65 keV. 
\begin{figure}[!t]
  \centering
  \includegraphics[width=9cm]{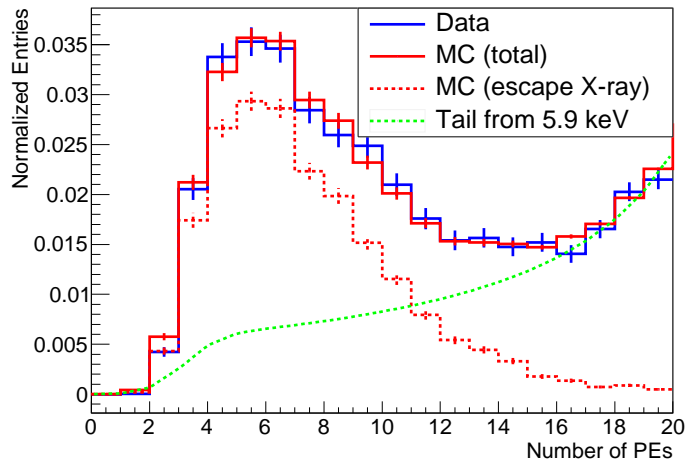}
  \caption{(Colour online) PE distribution for the escape peak. 
  Solid blue and red histograms represent the observed data and MC. 
  The dashed red histogram and the green line illustrate the best-fit result for the escape peak component and tail component from the 5.9 keV X ray.
  }
  \label{Fe_best}
\end{figure}
Figure \ref{Fe_best} shows the distribution of the number of PEs for the escape peak. 
The scintillation efficiency at 1.65 keV was evaluated by comparing these escape peaks in the data (solid blue line) and total MC (solid red line) considering systematic uncertainties
such as the source assembly with its shadowing and reflection effects, trigger efficiency, the choice of fitting functions. 
The dashed red histogram represents the PE distribution only for escape peaks,
whereas the green line represents the PE distribution} for tail component from 5.9 keV peak, which was caused by the shadowing effect of the calibration source.
The tail component was also modelled with parameters and simultaneously fitted because of the uncertainty.
Total MC distribution was then calculated as the summation of these two components.
Considering all the systematic and statistical uncertainties, the scintillation efficiency at 1.65 keV was estimated to be 39$_{-4}^{+4}$\% of that of 122 keV.
As the result of this calibration, the energy scale at 1.65 keV became ~20\% lower than the previous scale used in \cite{XMASS_MOD,XMASS_MOD2017}.
The energy threshold for sub-GeV DM analysis via bremsstrahlung was set to 1.0 keV$_{\mathrm{ee}}$, since the uncertainty below that energy considerably increases. 
The scintillation efficiency at 1 keV$_{\mathrm{ee}}$ was estimated to be $31^{+7}_{-4}$\% of that of 122 keV.

In addition to the scintillation efficiency, detector resolution was also calibrated using these peaks.
The resolution of the detector at 1.65 keV was estimated from the calibration measurement to be 40\% , and Gaussian smearing was applied to MC to reproduce the data.
This extra smearing was (17$\pm$10)\%.
The 10\% uncertainty was mainly due to the surface roughness and reflection of the source. 

The nonlinear response for nuclear recoil with energy over 3 keV$_{\rm{nr}}$ was estimated using the scintillation efficiency at zero electric field in \cite{XENON_Leff}.
The LUX group conducted a nuclear recoil calibration \cite{LUX_Leff} using neutrons from a deuterium-deuterium beam at 180 V/cm, the resultant scintillation efficiency for nuclear recoil is used to estimate the response for nuclear recoil energy below 3 keV$_{\rm{nr}}$.
The existence of an electric field in \cite{LUX_Leff} reduces the light yield.
The amount of the reduction due to electric field was considered to be level of 10\% \cite{ZEP,XENON_NR}.
Although the XMASS detector is operated under zero electric field, we used the unaltered results with 10\% uncertainty, a typical reduction amount.
The energy threshold for multi-GeV DM analysis via nuclear recoil is set to 2.3 keV$_{\rm{nr}}$ such that we could suppress an impact of the systematic error caused by the flasher events explained in Section 6, to be smaller than other errors.
At this energy, a 50\% trigger efficiency of the signal simulation (8 GeV) was obtained; this threshold corresponds to 2.3 PEs.
The scintillation efficiency at this energy was changed to 8.5\% from 6.5\%.

\section{Analysis and results for sub-GeV DM}

Event selection was applied in two stages that we referred to as standard and likelihood cuts \cite{XMASS_MOD2017}.
The standard cut eliminates events that are indicative of electric noise, afterpulses, or Cherenkov emissions inside the quartz window of PMTs rather than physical interactions in the detector.
Following the standard cut, we applied the likelihood cut on the basis of PE hit patterns, which removes background events occurring in front of a PMT window or near the detector wall. 

The treatment of systematic uncertainties was the same as in \cite{XMASS_MOD2017}. 
The dominant systematic uncertainty in this analysis was associated with the variation in the PE yield during exposure.
As discussed in section 4, the variation in the LXe absorption length causes a variation in the PE yield.
This variation both distorts the spectrum and changes the cut efficiency.
These effects were corrected based on the calculation of the relative change in the spectrum using MC simulations.
To correct for each time/energy bin of measured data, MC simulations with corresponding absorption lengths derived from $^{57}$Co calibration in each period were generated.
Using these simulation results, the correction factors for the corresponding time/energy bins were calculated.
These correction factors for each bins are referred to as the relative efficiency.

As it is explained in \cite{XMASS_FV}, the main source of background in these energy regions is $^{238}$U and $^{210}$Pb contained in the sealing material between the quartz window and metal body of each PMT.
Since the relative efficiency depends on the spectrum shape of the expected background, the uncertainties were evaluated by comparing reasonable background models. 
This uncertainty of the background contributed the most to the systematic error in the relative efficiency, 1.2\% and 2.5\% at 1 and 5 keV$_{\rm{ee}}$, respectively.
Note that these errors of the count rate have a correlation between each energy and time bin.
The next--leading contribution came from the gain instability in the waveform digitizers between April 2014 and September 2014.
During that period, a different calibration method was used for the digitizers.
This variation contributed an extra uncertainty of 0.3\% to the energy scale.
Other contributions from the uncertainty in the PMT gain calibration using a LED, trigger-threshold stability and timing calibration were negligible.

The dataset was divided into 86 time bins ($t_{{\rm bins}}$) with roughly 15 live days in each bin.
The data in each time bin was further divided into energy-bins ($E_{{\rm bins}}$) with bin width of 0.5~keV$_{\rm{ee}}$.
For the DM search through the bremsstrahlung effect, the data was fitted in the energy range from 1.0 to 20~keV$_{\rm{ee}}$.

Minimum-$\chi^{2}$ fitting was performed in the annual modulation analysis.
In this analysis, the `pull method' \cite{PULL}, one of the two different methods in previous analyses \cite{XMASS_MOD}, was used to fit all energy and time bins simultaneously and to treat the correlated errors.
The $\chi^{2}$ function is defined as follows:
\begin{equation}
\label{eq_chi}
\chi^2 = \sum\limits_{i}\limits^{E_{{\rm bins}}} \sum\limits_{j}\limits^{t_{{\rm bins}}} 
\left(\frac{(R^{{\rm data}}_{i,j}-R^{\rm ex}_{i,j}(\alpha,\beta))^2}{\sigma({\rm stat})^2_{i,j}+\sigma({\rm sys})^2_{i,j}}\right)+\alpha^{2}+\sum^{Nsys}_k \beta_k^{2}, 
\end{equation}
where $R_{i,j}^{\rm data}$, $R_{i,j}^{\rm ex}$, are the data and expected number of events for the $i$-th energy and $j$-th time bins after considering the efficiency of all event selections, respectively.
$\sigma(\rm{stat})$$_{i,j}$ and $\sigma(\rm{sys)}$$_{i,j}$ are the statistical and systematic uncertainty of the expected number of events, respectively. 
The `pull terms', $\alpha$ and $\beta_k$ represent the size of the systematic uncertainties that have correlations in energy bins or time bins.
$\alpha$ is overall size of the relative efficiency errors common for all energy bins.
Therefore, the error size of each bin changes simultaneously during the fit procedure.
$\alpha$ = 1 ($-$1) corresponds to the 1 $\sigma$ ($-$1 $\sigma$) correlated systematic error on the expected event rate. 
$\beta_k$ is the $k$-th systematic uncertainty of the signal simulation caused by the properties of LXe. 

The uncertainties for scintillation time constants and the scintillation efficiency for the electron-recoil signal were considered.
These uncertainties correlatively alter the signal spectrum between energy bins.
For time constants, two components referred to as fast and slow component were used on the basis of the $\gamma$-ray calibration of the XMASS-I detector \cite{DKconPaper}.
These were 2.2 and 27.8$^{+1.5}_{-1.0}$ ns, respectively, with the fast component fraction of 0.145$^{+0.022}_{-0.020}$.
For the scintillation efficiency, the uncertainty described in section 4 was used.
We assumed that the signal efficiency below 1.0 keV$_{\rm ee}$ is zero because of the uncertainty in the scintillation efficiency.
The effect of the uncertainty of the energy resolution is much smaller than that of scintillation efficiency and is negligible.

\begin{figure}[!t]
  \centering
  \includegraphics[width=8cm]{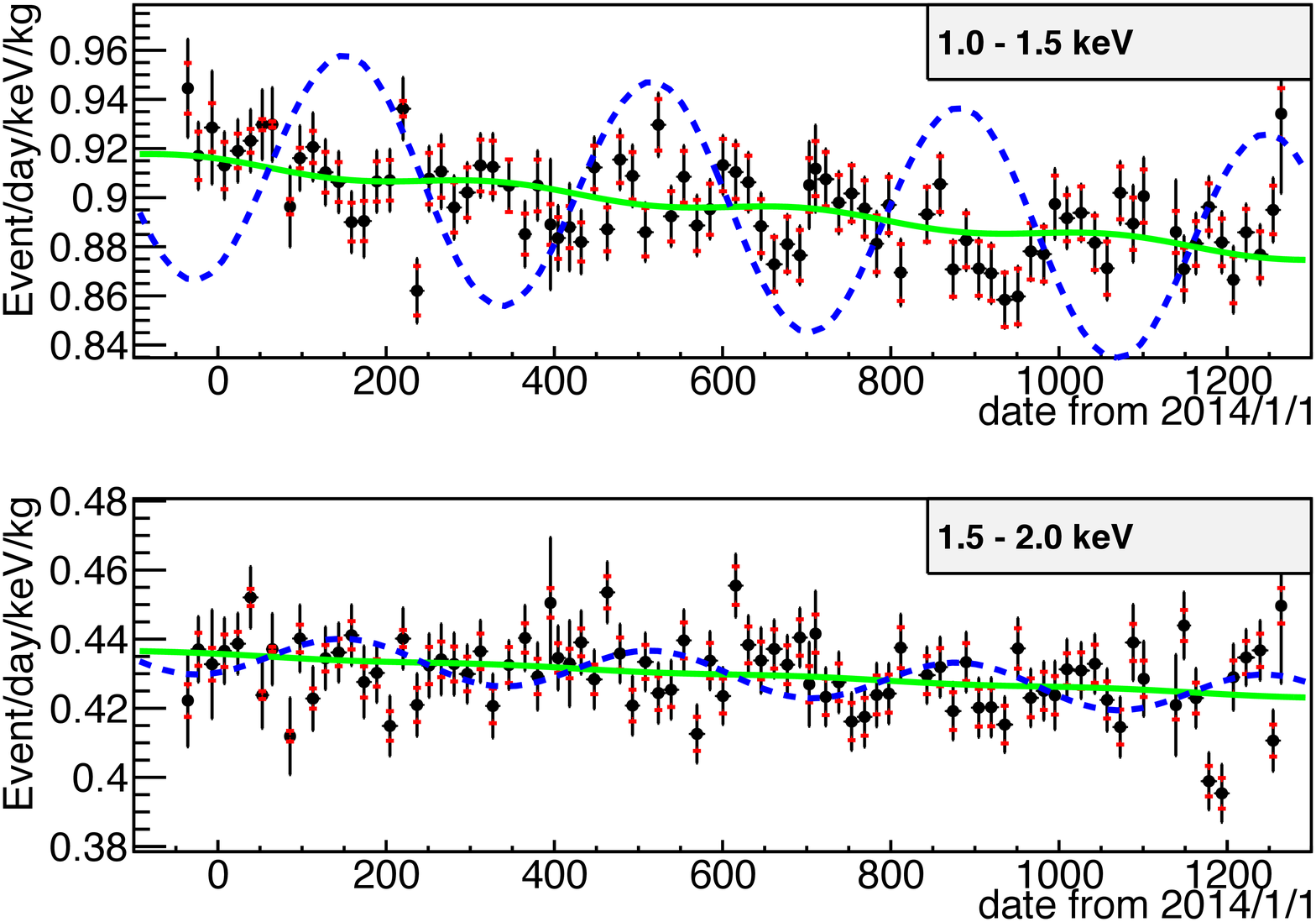}
  \caption{(Colour online) Result of best fit for data at 1.0--1.5 and 1.5--2.0 keV$_{\rm{ee}}$.
  The black points indicate data with the statistical uncertainty of the count rate. 
  The red brackets indicate the 1 $\sigma$ systematic error for each time bin.
  The green line indicates the best-fit result for the bremsstrahlung spectrum. 
  The blue dash line indicates the expected amplitude for 0.5 GeV DM at $3\times 10^{-32}$ cm$^2$ sensitivity.
  All data points and lines are corrected for the efficiency curve with the best-fit $\alpha$. }
  \label{Brems_data}
\end{figure}

The expected number of events $R^{\rm ex}_{i,j}(\alpha,\beta)$ is then expressed as follows: 
\begin{eqnarray}
 R_{i,j}^{\rm ex}(\alpha,\beta)&=& \int_{t_{j}-\frac{1}{2}\Delta t_{j}}^{t_{j}+\frac{1}{2} \Delta t_{j}} \biggl\{\epsilon^b_{i,j}(\alpha)\cdot (B^b_it+C^{b}_{i}) \nonumber \\
 &+& \sigma_{\chi n} \cdot \epsilon^s_{i,j}\cdot \Bigl[ C^{s}_{i}(\beta)+ A^{s}_{i}(\beta) \cos \big( 1\pi \frac{t-\phi}{T} \big) \Bigr] \biggr\} dt, \nonumber \\
\label{eq:MD}
\end{eqnarray}
where $t_{j}$ and $\Delta t_{j}$ are the center and width of the $j$-th time bin, respectively;
$\sigma_{\chi n}$ is the DM-nucleon cross section; $\epsilon^b_{i,j}(\alpha)$ and $\epsilon^s_{i,j}(\alpha)$ are the relative efficiencies for the background and signal, respectively.
To account for the changing background rates from long-lived isotopes, we added a simple linear function with slope $B^b_i$ and constant $C_i^b$ in the $i$-th bin.
The source of the decay was considered as $^{210}$Pb, which has a half-life of 22.3 years.
$A^s_i(\beta)$ represents the amplitude, and $C^{s}_i(\beta)$ represents the unmodulated component of the signal in the $i$-th energy bin.
In this analysis, the signal efficiencies for each DM mass were estimated using the MC simulations of uniformly injected photons from the bremsstrahlung effect in the LXe volume.
The unmodulated component and amplitude of the signal spectrum were calculated for a particular cross section and mass of DM.
The sub-GeV DM analysis was conducted for DM masses between 0.32 and 1.00 GeV.
Figure \ref{Brems_data} shows the observed event rate with the best fit and expected time valuation for 0.5 GeV at 1.0--1.5 and 1.5--2.0 keV$_{\rm{ee}}$.
The search for DM mass more than 1 GeV via this bremsstrahlung effect has not been performed because the assumptions for the signal calculation in \cite{subGeV}, such as that for form factor were not proper.
The deviation was $\sim$0.3\% and $\sim$3\% at a maximum momentum transfer of 1 and 3 GeV DM, respectively.
The best fitted cross section from the data was -1.4$^{+1.3}_{-1.6}$ $\times 10^{-33}$ cm$^2$ at 0.5 GeV. 
The best fitted $\chi^2$/NDF was 3333.8/3188, and pull parameter $\alpha$ was 0.6 for 0.5 GeV.

Considering that we found no significant signal, the 90\% confidence level (CL) upper limit on the DM-nucleon cross section
$\sigma_{up}$ was calculated by the Bayesian approach \cite{PDG}:
\begin{eqnarray}
  \int^{\sigma_{up}}_0 Pd\sigma_{\chi n} / \int^\infty_0 Pd\sigma_{\chi n} = 0.9, 
\label{eq:90CL}
\end{eqnarray}

where P is the probability function defined as follows:
\begin{eqnarray}
  {\rm P} = {\rm exp}\left(\frac{\chi^2(\sigma_{\chi n})-\chi^2_{min}}{2}\right).
\label{eq:Prob}
\end{eqnarray}

\begin{figure*}[h!]
  \centering
  \includegraphics[width=18cm,clip]{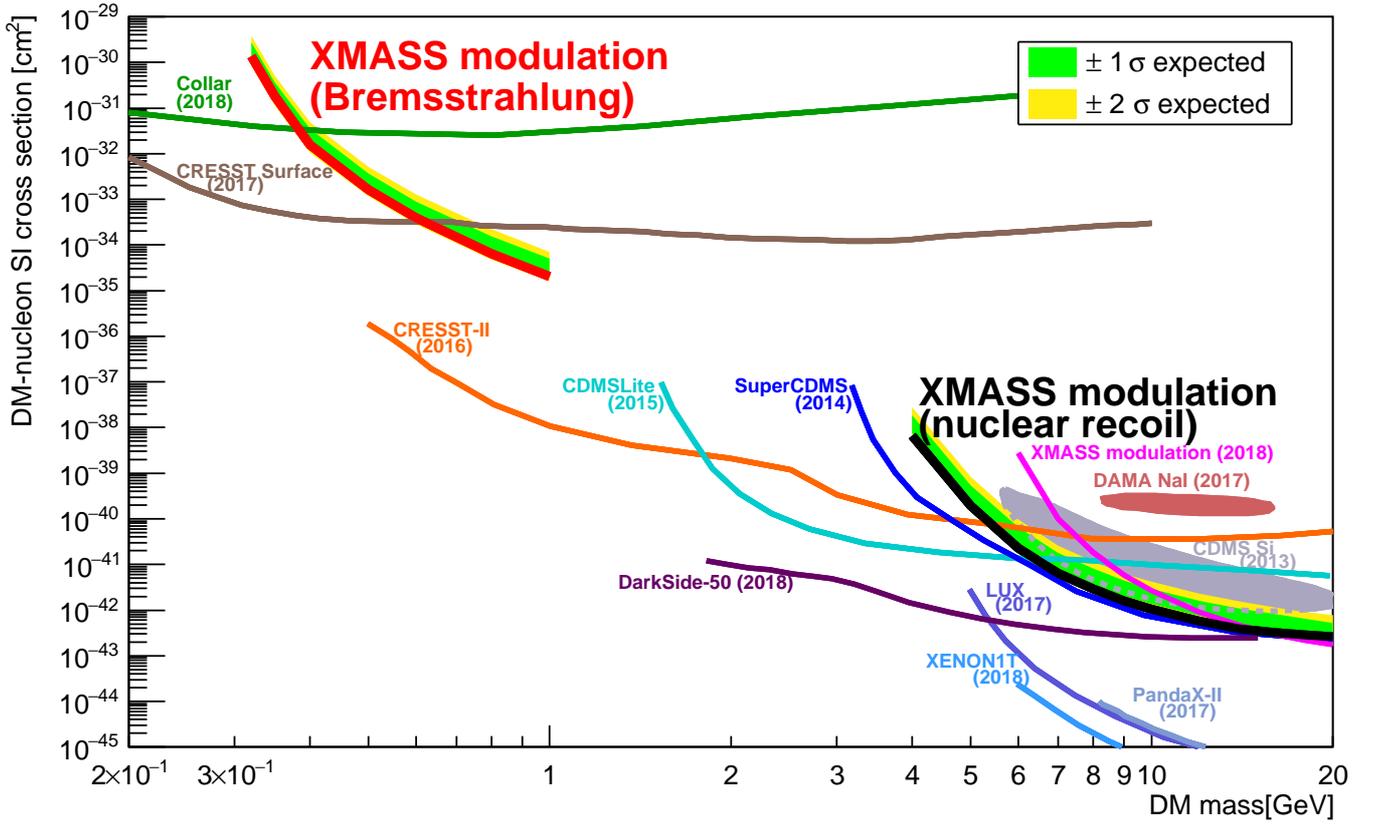}
  \caption{(Colour online)
  Summary of the search results. 
  The red line is the result of the bremsstrahlung analysis for 0.32--1 GeV DM.
  For comparison, data from the CRESST sapphire surface detector \cite{CRESSTSurPaper} and CRESST-II \cite{CRESSTIIPaper}, which are searching for the elastic nuclear recoil signals, are shown in each colour.
  The black line shows the result of the nuclear recoil search at 4--20 GeV.
  For comparison, results from CDMS-Si \cite{CDMSSiPaper}, CDMSLite \cite{CDMSLitePaper}, SuperCDMS \cite{SuperCDMSPaper}, LUX \cite{LUX}, XENON1T \cite{XENON1T}, PandaX-II \cite{PANDA}, DAMA/LIBRA \cite{DAMAPaper, DAMA_WIMP}, and XMASS-I \cite{XMASS_MOD2017}, DarkSide-50 \cite{DS}, and the liquid scintillator experiment by Collar \cite{JIC} are shown for each colour.
  The green and yellow bands for each result show the $\pm$ 1 $\sigma$ and $\pm$ 2 $\sigma$ expected sensitivity of 90\% CL upper limits for the null-amplitude case, respectively.
  }
  \label{Summary}
\end{figure*}

The result of the DM search via the bremsstrahlung effect is shown in the sub-GeV region of Fig.~\ref{Summary}.
The expected sensitivity for the null-amplitude case is calculated by using the statistical samples.
They were generated based on the event rate obtained from a fitted result of data with only background components decreasing linearly in time, as described in \cite{XMASS_MOD, XMASS_MOD2017}.
When generating these statistical samples, data for each period and each energy bin was fitted without the signal amplitude in the first step.
Thereafter, the expected number of events in each period was calculated while considering systematic errors such as relative efficiency. 
Finally, the Poisson fluctuation of the number of events was calculated for each energy bin, on the basis of the livetime of each period. 
One thousand sets of statistical samples were generated, and the 90\% CL upper limit sensitivity was calculated for each sample.
The 90\% CL sensitivity for DM at 0.5 GeV was 2.4$^{+1.2}_{-0.8}$ $\times 10^{-33}$ cm$^2$ (the range containing 68\% of statistical samples) and our upper limit was 1.63 $\times 10^{-33}$ cm$^2$ (p-value: 0.27).

\section{Analysis and results for multi-GeV DM}
An additional search for multi-GeV DM signals from elastic nuclear recoil was conducted.
The analysis was mostly identical to that of the sub-GeV DM search, but data on energy less then 1.0 keV$_{\rm{ee}}$ were analysed using nuclear recoils as low as 2.3 PE ($\sim$~2.3 keV$_{\rm{nr}}$, $\sim$~0.5 keV$_{\rm{ee}}$).
This type of data, the low threshold data, has been recorded since December 8, 2015 with three PMT hit trigger.
The total exposure of the data was 0.63 ton-years. The signal efficiency after all the data selection was improved from 5\% and 10\% to 10\% and 15\% at the lowest energy bin (2.3 -- 4.8 kev$_{\rm{nr}}$) for 4 GeV and 8 GeV DM, respectively.
This improvement of the trigger condition occurred due to the decrease of dark hits of each PMT.
Average dark hits for each PMT were approximately 15 Hz at earlier periods and decreased to approximately 5 Hz during the operation.
After the several data-taking tests, we were able to record stable data with the three-PMT hit triggers.

The primary uncertainty in the low-threshold data came from a weak light emission of the PMTs with a one PE. 
From the measurement for several PMTs in room temperature,
the probability of the emission per a one PE was $\sim$0.3 - 1.0\%.
Given that the light emission occurs even after dark hits, changes in the dark hits for each PMT directly change the event rate around the threshold.
Thus, an additional condition for the run selection was applied to suppress this uncertainty; periods where the dark-hit rates for individual PMTs as well as the total dark-hit rate among all the PMTs changed more than 500 Hz from the nominal values were removed from the analysis.
Furthermore, the event with this light emission has characteristic timing and angular distributions of hit PMTs; the time difference between the PMT emitting the light and other PMTs receiving the light after emission distributed more than 35 ns and the latter PMTs were located within 50 degrees from the former PMT.
Therefore, if any pair of hits in the events agrees with these conditions, the event was eliminated from the analysis.
This event selection, referred to as a flasher cut, was applied only for three PMT hit events, and the uncertainty due to the weak flash effect after this cut is 0.4\% at maximum.


The $\chi^2$ and expected event rate functions for the time variation fitting are the same as those in the sub-GeV DM analysis except for the energy range. 
Most of the uncertainty for elastic nuclear recoil signal is discussed in \cite{XMASS_MOD2017}, only the uncertainty of the xenon scintillation efficiency for nuclear recoil is different.
As discussed above in section 4, the measurements for energy below 3 keV$_{\rm{nr}}$ in \cite{LUX_Leff} are considered.

From the multi-GeV DM analysis, we obtained the best-fit cross section between 4 and 20 GeV DM mass.
The best-fit cross section is -3.8$^{+2.0}_{-4.5}$ $\times$ 10$^{-42}$ cm$^2$ at 8 GeV, and no significant signal was found in this analysis including other mass.
Because of this, a 90\% CL upper limit on the DM-nucleon cross section was determined.
The 90\% CL sensitivity at 8 GeV was 5.4$^{+2.7}_{-1.7}$ $\times$ 10$^{-42}$ cm$^2$, and the upper limit was 2.9 $\times$ 10$^{-42}$ cm$^2$ (p-value: 0.11).
The result of the DM search via the nuclear recoil signal is plotted in the multi-GeV region of Fig.~\ref{Summary}. 
The upper limits and allowed regions determined by other experiments are also shown. 

Compared with the result from the previous analysis of XMASS data \cite{XMASS_MOD2017}, the result of the present analysis is approximately 6.7 times better at 8 GeV.
Because both the low-threshold data and the new scintillation efficiency below 3 keV$_{\rm{nr}}$in \cite{LUX_Leff} improve the sensitivity.
The search for DM mass below 3 GeV was not performed via nuclear recoil.
This is because the maximum recoil energy is below 1 keV$_{\rm{nr}}$, which is the lowest calibrated energy in \cite{LUX_Leff}.

\section{Conclusion}
We carried out the annual modulation analysis for XMASS-I data to search for the sub-GeV and multi-GeV DM via the bremsstrahlung effect and elastic nuclear recoil, respectively.
The former search limits the parameter space of DM with a mass of 0.5 GeV to below 1.6 $\times$ $10^{-33}$ cm$^2$ at 90\% CL.
This is the first experimental result for a sub-GeV DM search focused on annual modulation and bremsstrahlung photons emitted by inelastic nuclear recoils.
The additional search for the multi-GeV DM with the lower threshold data obtained a limit for the parameter space of DM with a mass of 8 GeV to below 2.9 $\times$ $10^{-42}$ cm $^2$ at 90\% CL.

\section*{Acknowledgements}

We gratefully acknowledge the cooperation of Kamioka Mining and Smelting Company. 
This work was supported by the Japanese Ministry of Education,
Culture, Sports, Science and Technology,
the joint research program of the Institute for Cosmic Ray Research (ICRR),
the University of Tokyo,
Grant-in-Aid for Scientific Research, 
JSPS KAKENHI Grant Number, 19GS0204, 26104004, and partially
by the National Research Foundation of Korea Grant funded
by the Korean Government (NRF-2011-220-C00006).


\newpage
\bibliography{biblio} 

\begin{thebibliography}{10}
\expandafter\ifx\csname url\endcsname\relax
  \def\url#1{\texttt{#1}}\fi
\expandafter\ifx\csname urlprefix\endcsname\relax\def\urlprefix{URL }\fi
\expandafter\ifx\csname href\endcsname\relax
  \def\href#1#2{#2} \def\path#1{#1}\fi

\bibitem{WIMPs}
M.~W. Goodman, E.~Witten\hspace{0pt}, Phys. Rev. D 31 (1985) 3059.

\bibitem{XENON1T}
E.~Aprile {\it et~al}. (XENON Collaboration), Phys. Rev. Lett. 121 (2018)
  111302.

\bibitem{LUX}
D.~S. Akerib {\it et~al}. (LUX Collaboration), Phys. Rev. Lett. 118 (2017)
  021303.

\bibitem{PANDA}
X.~Cui {\it et~al}. (PandaX-II Collaboration), Phys. Rev. Lett. 119 (2017)
  181302.

\bibitem{DUAN2018296}
G.~H. Duan {\it et~al}.\hspace{0pt}, Physics Letters B 778 (2018) 296.

\bibitem{AsymmDM1}
D.~B. Kaplan\hspace{0pt}, Phys. Rev. Lett. 68 (1992) 741.

\bibitem{AsymmDM2}
D.~E. Kaplan, M.~A. Luty, K.~M. Zurek\hspace{0pt}, Phys. Rev. D 79 (2009)
  115016.

\bibitem{AsymmDM3}
K.~PETRAKI, R.~R. VOLKAS\hspace{0pt}, International Journal of Modern Physics A
  28~(19) (2013) 1330028.

\bibitem{hiddenDM}
J.~L. Feng, J.~Kumar\hspace{0pt}, Phys. Rev. Lett. 101 (2008) 231301.

\bibitem{CDMSLitePaper}
R.~Agnese {\it et~al}. (SuperCDMS Collaboration), Phys. Rev. Lett. 116 (2016)
  071301.

\bibitem{CRESSTSurPaper}
G.~Angloher {\it et~al}. (CRESST Collaboration), Eur. Phys. J. C 77 (2017) 637.

\bibitem{DM_elec}
H.~An {\it et~al}.\hspace{0pt}, Phys. Rev. Lett. 120 (2018) 141801.

\bibitem{DM_elec_SLAC}
R.~Essig, J.~Mardon, T.~Volansky\hspace{0pt}, Phys. Rev. D 85 (2012) 076007.

\bibitem{subGeV}
C.~Kouvaris, J.~Pradler\hspace{0pt}, Phys. Rev. Lett. 118 (2017) 031803.

\bibitem{subGeV2}
C.~McCabe\hspace{0pt}, Phys. Rev. D 96 (2017) 043010.

\bibitem{Migdal}
M.~Ibe {\it et~al}.\hspace{0pt}, Journal of High Energy Physics 2018~(03)
  (2018) 194.

\bibitem{XMASS_MOD}
K.~Abe {\it et~al}. (XMASS Collaboration), Phys. Lett. B 759 (2016) 272 -- 276.

\bibitem{XMASS_MOD2017}
K.~Abe {\it et~al}. (XMASS Collaboration), Phys. Rev. D 97 (2018) 102006.

\bibitem{PDG}
C.~Patrignani {\it et~al}. (Particle Data Group), Phys. Rev. D 98 (2018) 530.

\bibitem{Lewin}
J.~Lewin, P.~Smith\hspace{0pt}, Astropart. Phys. 6 (1996) 87 -- 112.

\bibitem{MW}
M.~C. Smith {\it et~al}.\hspace{0pt}, Mon. Not. R. Astron. Soc. 379 (2007)
  755--772.

\bibitem{XMASS_Det}
K.~Abe {\it et~al}. (XMASS Collaboration), Nucl. Instr. Meth. A 716 (2013) 78
  -- 85.

\bibitem{DoublePE}
C.~Faham {\it et~al}.\hspace{0pt}, Journal of Instrumentation 10 (2015) P09010.

\bibitem{XMASS_Cal}
N.~Kim {\it et~al}. (XMASS Collaboration), Nucl. Instr. Meth. A 784 (2015) 499
  -- 503.

\bibitem{DOKE}
T.Doke, R.~Sawada, H.~Tawara\hspace{0pt}, River Edge, USA: World Scientific
  (2002) (2002) 17.

\bibitem{XENON_Leff}
E.~Aprile {\it et~al}. (XENON100 Collaboration), Phys. Rev. Lett. 107 (2011)
  131302.

\bibitem{LUX_Leff}
D.~S. Akerib {\it et~al.} (LUX Collaboration), arXiv:1608.05381.

\bibitem{ZEP}
M.~Horn {\it et~al}.\hspace{0pt}, Physics Letters B 705 (2011) 471.

\bibitem{XENON_NR}
E.~Aprile {\it et~al}. (XENON100 Collaboration), Phys. Rev. D 88 (2013) 012006.

\bibitem{XMASS_FV}
K.~Abe {\it et~al.} (XMASS Collaboration), arXiv:1804.02180, to be published by
  Physics~Letters~B.

\bibitem{PULL}
G.~L. Fogli {\it et~al}.\hspace{0pt}, Phys. Rev. D 66 (2002) 053010.

\bibitem{DKconPaper}
H.~Takiya {\it et~al}. (XMASS Collaboration), Nucl. Instr. Meth. A 834 (2016)
  192 -- 196.

\bibitem{CRESSTIIPaper}
G.~Angloher {\it et~al}. (CRESST Collaboration), Eur. Phys. J. C 77 (2017) 637.

\bibitem{CDMSSiPaper}
R.~Agnese {\it et~al}. (CDMS Collaboration), Phys. Rev. Lett. 111 (2013)
  251301.

\bibitem{SuperCDMSPaper}
R.~Agnese {\it et~al}. (SuperCDMS Collaboration), Phys. Rev. Lett. 112 (2014)
  241302.

\bibitem{DAMAPaper}
R.~Bernabei {\it et~al}.\hspace{0pt}, Eur. Phys. J. C 73 (2013) 2648.

\bibitem{DAMA_WIMP}
J.~Kopp, T.~Schwetz, J.~Zupan\hspace{0pt}, JCAP 2012 (2012) 001.

\bibitem{DS}
P.~Agnes {\it et~al}. (DarkSide Collaboration), Phys. Rev. Lett. 121 (2018)
  081307.

\bibitem{JIC}
J.~I. Collar\hspace{0pt}, Phys. Rev. D 98 (2018) 023005.

\end{thebibliography}
\bibliographystyle{elsarticle-num2}

\end{document}